\documentstyle[preprint,aps]{revtex}
\begin{document}
\draft
\preprint{\begin{tabular}{l}
\hbox to\hsize{January 17, 1997 \hfill SNUTP 96-073}\\[-3mm]
\hbox to\hsize{\hfill hep-ph/9701xxx }\\[5mm] \end{tabular} }
\bigskip
\title{ More on $R-$parity and lepton--family number violating 
\\ couplings from  muon(ium) conversion, and $\tau$ and $\pi^0$ decays }
\author{Jihn E. Kim$^{a,b}$ \thanks{jekim@phyp.snu.ac.kr}, 
Pyungwon Ko$^{c}$ \thanks{pko@phyb.snu.ac.kr},
and  Dae-Gyu Lee$^a$ \thanks{leedaegy@ctp.snu.ac.kr}}
\address{
$^a$ Department of Physics, Seoul National University,
Seoul 151-742, Korea        \\
$^b$ Center for Theoretical Physics, Seoul National University, 
Seoul 151-742, Korea        \\
$^c$ Department of Physics, Hong-Ik University, Seoul 121-791, Korea \\
}
\maketitle
\begin{abstract}
We present a new class of constraints to the lepton--family number and 
$R-$parity violating couplings from muonium conversion, 
$\mu^{-} + {_{22}^{48}}{\rm Ti} \rightarrow e^{-} 
+ {_{22}^{48}}{\rm Ti}$, 
a class of tau decays, $\tau \rightarrow l + $ (light meson) 
with $l= \mu~{\rm or}~ e$, and $J/\psi$ and $\pi^0$ decays into a lepton
pair.  We find that
$\mu^{-} + {_{22}^{48}}{\rm Ti} \rightarrow e^- + {_{22}^{48}}{\rm Ti}$
provides one of the strongest constraints along with $\Delta m_{K},
\Delta m_{B}$, $\mu \rightarrow e \gamma$ and the neutrinoless double 
$\beta$ decay.
Search for these lepton--family number violating (LFNV) decays 
forbidden in the standard model
is clearly warranted in various low-energy experiments such as Tau--Charm
Factories and PSI, {\it etc.}.
\end{abstract}


\narrowtext
\tighten

 \section{Introduction}
 \label{sec:one}

Lepton--family numbers are accidental global symmetries of the standard 
model (SM), and thus the electron, muon, and tau lepton numbers 
(denoted by $L_e, L_{\mu},$ and $L_{\tau}$, respectively) 
are separately conserved as
well as the total lepton number, $L_{tot} = L_e + L_{\mu} + L_{\tau}$.  
On the contrary, this is no longer true in the 
Minimal Supersymmetric Standard Model (MSSM) \cite{mssm}.  
Supersymmetry,  gauge invariance, and renormalizability 
do not forbid the following lepton number and/or baryon number violating 
terms in the renormalizable superpotential \cite{rp} :
\begin{equation}
W_{\not{R_{p}} } = {1 \over 2}~\lambda_{ijk}~L_{i}L_{j} E_{k}^{c}
+ \lambda_{ijk}^{'}~L_{i} Q_{j} D_{k}^{c}
+ {1 \over 2}~\lambda_{ijk}^{''}~U_{i}^{c} D_{j}^{c} D_{k}^{c} 
+ \mu_{i} L_{i} H_{2},
\end{equation}
where the meaning of $L$, $E^{c}$, $Q$, $D^{c}$, $U^{c}$, and $H_{2}$
should be self-evident, and 
the indices $i$, $j$, and $k$ refer to families.
The $SU(3)_c$ color and the $SU(2)_L$ group indices
are suppressed for simplicity, and we have 
$\lambda_{ijk} = - \lambda_{jik}$ and $\lambda_{ijk}^{''} =
- \lambda_{ikj}^{''}$. 
The first two and fourth terms in Eq.~(2) are lepton number violating, 
whereas the third term is baryon number violating. 
It has been well-known that
there is a very tight constraint on $\lambda^{'} 
\lambda^{''}$ from nonobservation of proton decay \cite{protondecay} 
\cite{chun}.

The most popular solution to such a stringent bound is to introduce
a discrete symmetry called $R-$parity defined as  
\begin{equation}
R_p \equiv (-1)^{3B+L_{tot}+2S},
\end{equation}
where $B$, $L_{tot}$, and $S$ are the baryon number, total lepton number, 
and intrinsic spin of a particle, respectively.  
Then the ordinary particles appearing in the SM as well 
as the extra  Higgs boson in the MSSM are 
$R-$parity even, whereas their superpartners are $R-$parity odd.
Therefore, the $R-$parity conservation implies that the superpartners of 
ordinary particles be always produced in pairs, 
and that the lightest supersymmetric particle (LSP) be stable.  
This property of LSP puts a strong constraint  
on the possible phenomenology at colliders. Also the LSP plays a potentially
important role in cosmology as a (cold) dark matter candidate \cite{coldDM}.
This interesting symmetry, the $R-$parity, 
can be introduced even naturally \cite{autoR},
that is, without any other symmetry 
except a gauge symmetry and supersymmetry. 

However, the existence of the $R-$parity symmetry itself has not been 
confirmed.
It is clearly worth looking for the $R-$parity violating
processes and deriving the constraints on the $R-$parity violating couplings.
The proton decay originated from the $R$-parity violating terms
can be evaded  by assuming a weaker condition 
than the $R$-parity conservation,
either $\lambda^{'} = 0$ or $\lambda^{''} = 0$.
The latter corresponds to the baryon-number conservation. 
The last term in Eq.~(2) can generate neutrino masses \cite{suzuki}, and 
have interesting phenomenological consequences. However, it 
is irrelevant to the four-fermion processes considered in this paper,
and thus will be ignored from now on.
In the case of the lepton-number conservation ($\lambda=\lambda^{'}=0$),
constraints
on the baryon number violating couplings $\lambda_{ijk}^{''}$ can be
obtained from various hadronic processes \cite{baryon}.
In this work, we relax the $R-$parity conservation
assuming the baryon-number conservation, $\lambda^{''} = 0$, 
and derive new bounds on $\lambda^{(')} \lambda^{(')}$.
There are many earlier papers where constraints on $\lambda^{(')}
\lambda^{(')}$ (assuming $\lambda^{''} = 0$) were derived from various
low-energy processes \cite{han}, including the neutrinoless double beta
decay \cite{mohapatra}.
Recently, Choudhury et al. \cite{roy} assumed that $\lambda^{''} = 0$, 
and obtained constraints on the lepton number violating terms, 
$\lambda^{(')} \lambda^{({'})}$, 
considering the neutral meson mixing,
the flavor changing decays of $K, B$ mesons, and 
rare three-body leptonic decays of $\mu$ and $\tau$ such as
$\mu \rightarrow 3 e$ and
$\tau \rightarrow 3 e, 3 \mu, e 2 \mu$, or $\mu 2 e$. 
They got quite stringent limits on some
combinations of these couplings and the masses of superpartners of ordinary
matter.  Still,  some couplings  remain either unconstrained
(such as $\lambda_{322}^{'}, \lambda_{323}^{'}$),
or only weakly constrained (such as $\lambda_{22k}^{'}, \lambda_{13k}^{'}$ 
except $\lambda_{133}^{'}$) from the consideration of Ref.~\cite{roy}.
So far the most stringent limits have 
come from  $\Delta m_{K}, \Delta m_{B},
K \rightarrow \mu^+ \mu^-, \mu \rightarrow e \gamma$  \cite{han,roy}
and the neutrinoless double beta decay experiments \cite{mohapatra},
all of which yield $\lambda^{(')} \lambda^{(')} < 10^{-6}-10^{-8}$.

In this work, we consider various low energy processes with lepton--family 
number violations (LFNV) which can be 
induced/affected by the  $\lambda$ and $\lambda^{'}$ couplings in Eq.~(1).
In Sec.~II, we consider the muonium 
$(M) \rightarrow$ antimuonium ($\overline{M}$) conversion 
and $\mu^{-}  + {_{22}^{48}}{\rm Ti} \rightarrow e^{-} {_{22}^{48}}{\rm Ti}$,
and find that the latter process gives one of the most stringent limit on 
$\lambda \lambda^{(')}$.  
In Sec.~III, we consider a class of 
$\tau$ decays with LFNV, $\tau \rightarrow l +$ (light meson),  
Here,  $l = e$ or $\mu$,  and the ``light meson'' represents a pseudoscalar 
($PS$) such as $\pi^0, \eta, K^{0}$, or a vector meson ($V$) 
such as $\rho^0, K^{*0}, \phi, \omega$. 
In Sec. ~IV, we derive  constraints from 
$J/\psi ({\rm or} \pi^0) \rightarrow \mu^{\pm} e^{\mp}$ and $\pi^0 \rightarrow 
e^+ e^-$. Then, we briefly summarize our results in Sec. V.
 
Before closing this section, let us write the $R-$parity violating interaction
lagrangian in terms of component fields :
\begin{eqnarray}
{\cal L}_{int, \not{R_{p}}} & = & 
\lambda_{ijk} ~\left[ ~\tilde{\nu}_{iL} \overline{e_{kR}} e_{jL} + 
\tilde{e}_{jL} \overline{e_{kR}} \nu_{iL} + \tilde{e}_{kR}^{*} 
\overline{(\nu_{iL})^c} e_{jL} ~\right]
\nonumber \\
& + & \lambda_{ijk}^{'}~\left[ ~ \left( 
\tilde{\nu}_{iL} \overline{d_{kR}} d_{jL}
+ \tilde{d}_{jL} \overline{d_{kR}}  \nu_{iL} + \tilde{d}_{kR}^{*} 
\overline{(\nu_{iL})^c} d_{jL} \right)  \right.
\label{eq:int}
\\
& & \left. - V_{jp}^{\dagger} \left(~ \tilde{e}_{iL} \overline{d_{kR}} u_{pL}
+ \tilde{u}_{pL} \overline{d_{kR}} e_{iL} + \tilde{d}_{kR}^{*}
 \overline{(e_{iL})^c} u_{pL}~ \right) ~\right] + {\rm  h.c.}.   \nonumber
\end{eqnarray}
We have taken into account the flavor-mixing effects in the 
up-quark sector in terms of the CKM matrix elements, $V_{jp}$.  
The disalignment between fermion and sfermion fields will be ignored,
since it is strongly constrained from the suppression of the Flavor Changing 
Neutral Current (FCNC) processes.  The sparticle fields in Eq.~(3) 
are assumed to be the mass eigenstates.

Integrating out the superparticles such as sneutrinos or $u-$squarks, we get 
the effective lagrangian involving four fermions in the SM.
(In this work, we will not be concerned about the four-fermion
 interactions with neutrinos such as $\pi \rightarrow l \nu$.)
For example, by integrating out the sneutrino fields, we get the 
$| \Delta S | = 2$ effective lagrangian,  
\begin{equation}
{\cal L}_{eff}^{| \Delta S | = 2} = - \sum_{n} 
{\lambda_{n21}^{'} \lambda_{n12}^{'*}
\over m_{\tilde{\nu}_n}^2 }~\overline{d_R} s_{L}~\overline{d_L} s_{R},
\end{equation}
and similarly for the $| \Delta B | = 2$ effective lagrangian.
One can also get the effective lagrangian for $q_{i} + \overline{q_j} 
\rightarrow
e_{k} + \overline{e}_{l}$ by integrating out the sneutrino and the 
squark fields. The resulting effective lagrangian contributes to the 
processes, $\mu + {_{22}^{48}{\rm Ti}} \rightarrow e + {_{22}^{48}{\rm Ti}}$
and $\tau \rightarrow l + PS ({\rm or}~V)$, where $l = e$ or $\mu$, 
$PS = \pi^{0}, \eta,$ or $K$, and $V = \rho^{0},\omega,K^{*0},$ or $\phi$.
For $q = d$, we have \footnote{We do not agree with D.
Choudhury and P. Roy \cite{roy},
in the detailed form of the effective lagrangian for
$d_{i} + \overline{d_j} \rightarrow e_{k} + \overline{e}_{l}$.
Compare our Eq.~(5) with Eq.~(7) of Ref.~\cite{roy}.}
\begin{eqnarray}
{\cal L}_{eff} (d_{i} + \overline{d_j} \rightarrow e_{k} + \overline{e_l} )
& = &  - \sum_{n} {1 \over m_{\tilde{\nu}_{Ln}}^2}~ \left[ 
\lambda_{nij}^{'} \lambda_{nkl}^{*} \overline{e_{kL}} e_{lR}
\overline{d_{jR}} d_{iL}
+  \lambda_{nlk} \lambda_{nji}^{'*} \overline{e_{kR}} e_{lL}
\overline{d_{jL}} d_{iR}
\right]
\nonumber \\
& + & \sum_{m,n,p}~{V_{np}^{\dagger} V_{pm} \over 2 m_{\tilde{u}_{Lp}}^2}~
\lambda_{lnj}^{'} \lambda_{kmi}^{'*} \overline{e_{kL}} \gamma^{\mu} e_{lL}
\overline{d_{jR}} \gamma_{\mu} d_{iR}.
\end{eqnarray}
The first term comes from the sneutrino exchanges, whereas the second comes
from the $u-$squark exchanges.  We have used the Fierz transformation in 
order to get the second term.    
There is another effective lagrangian for $q_{i} + \overline{q_j} \rightarrow 
e_{k} + \overline{e}_{l}$, with $q $'s being up-type quarks, 
which can be obtained from Eq.~(3) by integrating out the $d-$squark fields :
\begin{eqnarray}
{\cal L}_{eff} ( u_{i} + \overline{u_j} \rightarrow e_{k} + \overline{e}_{l} ) 
& = & - \sum_{m,n,p}  \lambda_{lmp}^{'} \lambda_{knp}^{'*}
{V_{mi}^{\dagger} V_{jn}
 \over m_{\tilde{d}_{Rp}}^2}~\overline{\left( e_{lL} \right)^c} 
u_{iL} \overline{u_{jL}} \left( e_{kL} \right)^c 
\\   
& \rightarrow &  -  \sum_{m,n,p}  \lambda_{lmp}^{'} \lambda_{knp}^{'*}
{V_{mi}^{\dagger} V_{jn}
\over 2 m_{\tilde{d}_{Rp}}^2}~\overline{e_{kL}} \gamma^{\mu} e_{lL}
~\overline{u_{jL}} \gamma_{\mu} u_{iL},
\end{eqnarray}
after the Fierz transformation.

\section{Constraints from muon conversion}
\label{sec:two}

\subsection{Muonium $\rightarrow$ antimuonium conversion}
\label{subsec:muonium}

Let us first consider the muonium conversion,  $M(\equiv \mu^+ e^-) 
\rightarrow \overline{M}(\equiv \mu^- e^+)$.
The four--lepton effective  lagrangian relevant 
to the muonium conversion ($\Delta L_{\mu} = - \Delta L_{e} = -2$) 
can be obtained from Eq.~(\ref{eq:int}) 
by integrating out the sneutrino fields :
\begin{equation}
{\cal L}_{eff} (\mu^+ e^- \rightarrow \mu^- e^+) = -{\lambda_{321} 
\lambda_{312}^{*} \over m_{\tilde{\nu}_{L3}}^2 }~
\overline{\mu_R} e_{L} \overline{\mu_L} e_R,
\label{eq:muonium}
\end{equation}
In Eq.~(\ref{eq:muonium}), we have used the antisymmetry of the couplings, 
$\lambda_{ijk} = - \lambda_{jik}$,
in  order to simplify the sneutrino contributions.
The muonium conversion probability is usually translated into the upper 
limit on the hypothetical coupling $G_{M\overline{M}}$ defined as
\begin{equation}
{\cal L} (M \rightarrow \overline{M}) = {G_{M\overline{M}} \over \sqrt{2}}~
\left( \bar{\mu} e \right)_{V-A}  \left( \bar{\mu} e \right)_{V+A}
+ {\rm h.c.}.
\label{eq:v-a}
\end{equation}
Our effective Lagrangian, Eq.~(8),  is the same as 
Eq.~(\ref{eq:v-a}) after the 
Fierz transformation, with the following identification
\begin{equation}
{ G_{M\overline{M}} \over \sqrt{2}} = {\lambda_{321}
\lambda_{312}^{*} \over 8 m_{\tilde{\nu}_{L3}}^2 }.
\end{equation}
Therefore,  the conventional limit on $ G_{M\overline{M}}$ can be readily 
translated into the $R_p-$violating couplings.

The muonium conversion probability depends on the external magnetic 
field $B_{\rm ext}$ in a nontrivial way. This subject was recently 
addressed in detail by a few groups \cite{hou}, and we use their results 
in the following.  From the present upper limit on the transition probability
for the external magnetic field $B_{\rm ext} = 1.6$ kG,
\begin{equation}
P_{exp} (M \rightarrow \overline{M}) < 2.1 \times 10^{-9}  ~~~~~(90 \% C.L.),
\end{equation}
one gets the following contraint on $G_{M\overline{M}} < 9.6 \times 10^{-3}
G_F$ \footnote{  This type of interaction also arises in theories
with dilepton-gauge bosons ($Y^{\pm}, Y^{\pm \pm}$) \cite{sasaki},  such as
the $331$ model  considered by Frampton {\it et al.} \cite{frampton}.
This limit on the coupling $G_{M \bar{M}} $  is translated into a lower
bound on the mass of the dilepton-gauge boson, $M_{Y^{\pm \pm}}^2 >
(690 ~{\rm GeV})^2$.}.
This in turn implies that
\begin{equation}
| \lambda_{231} \lambda_{132}^{*} | < 6.3 \times 10^{-3}~
\left( m_{\tilde{\nu}_{L3}} \over 100~{\rm GeV} \right)^2.
\end{equation}
This constraint on the $R-$parity violating $\lambda$ 
couplings is in the same order with other constraints derived from 
lepton-flavor violating $\tau$ decays such as 
$\tau \rightarrow 3 l$ or $l l^{'+} l^{'-}$ 
(with $l, l^{'} = \mu$, or $ e$) \cite{roy}. 


\subsection{$\mu^{-} + {_{22}^{48}}{\rm Ti}\rightarrow e^{-}
+ {_{22}^{48}}{\rm Ti}$ conversion}
\label{subsec:ti}
 
In this subsection, let us consider the $\mu^{-} + {_{22}^{48}}{\rm Ti} 
\rightarrow e^{-} 
+ {_{22}^{48}}{\rm Ti}$ induced by
the $R-$parity violating $\lambda^{'} \times \lambda^{({'})}$ terms.
The relevant effective lagrangian at the parton level can be written as 
\begin{eqnarray}
{\cal L}_{eff} & = & {1 \over 2}~
\overline{e_L} \gamma_{\alpha} \mu_{L}~
\left[ A_{\mu {\rm Ti}}^{d}~\overline{d_R} \gamma_{\alpha} d_{R} +
A_{\mu {\rm Ti}}^{u} ~\overline{u_L} \gamma_{\alpha} u_{L} \right]
 \nonumber   \\
& + &  {1 \over 2}~
\left[~ S_{\mu {\rm Ti}}^{d,1}~ \overline{e_L} \mu_{R} ~
\overline{d_R} d_{L} + S_{\mu {\rm Ti}}^{d,2}~ \overline{e_R} \mu_L ~
\overline{d_L} d_R ~ \right],
\end{eqnarray}
where $A_{\mu {\rm Ti}}^{u,d}$ and $S_{\mu {\rm Ti}}^{d}$ 
can be obtained from Eqs.~(5) and (6) as follows :
\begin{eqnarray}
A_{\mu {\rm Ti}}^{d} & = & + \sum_{m,n,p} {V_{np}^{\dagger} V_{pm} \over  
m_{\tilde{u}_{Lp}}^2}~ \lambda_{2n1}^{'} \lambda_{1m1}^{'*},
\\
& \rightarrow & + \sum_{n} {\lambda_{2n1}^{'} \lambda_{1n1}^{'*} \over
m_{\tilde{u}_{Ln}}^2} ~~~~~{\rm for}~ V_{np} = \delta_{np},
\nonumber   
\\
A_{\mu {\rm Ti}}^{u} & = & -  \sum_{m,n,p} {V^{\dagger}_{m1} V_{1n} \over
m_{\tilde{d}_{Rp}}^2}~\lambda_{2mp}^{'} \lambda_{1np}^{'*},
\\
& \rightarrow & - \sum_{n} {\lambda_{21n}^{'} \lambda_{11n}^{'*} \over
m_{\tilde{d}_{Rn}}^2} ~~~~~{\rm for}~ V_{np} = \delta_{np},
\nonumber   
\\
S_{\mu {\rm Ti}}^{d,1} & = &  - \sum_{n} {2 \over m_{\tilde{\nu}_{L,n}}^2}~
\lambda_{n11}^{'} \lambda_{n12}^{*},
\\
S_{\mu {\rm Ti}}^{d,2} & = &  - \sum_{n} {2 \over m_{\tilde{\nu}_{L,n}}^2}~
\lambda_{n11}^{'*} \lambda_{n21}. 
\end{eqnarray}
In many supersymmetric theories with lepton--family number violation, 
the $\mu^{-} \rightarrow e^{-}$
conversion on the ${_{22}^{48}}{\rm Ti}$ nucleus occurs through 
the electroweak penguin 
diagram, $\mu^{-} \rightarrow e^{-} + \gamma^{*} ({\rm or}
~Z^{*})$, or through the box diagrams, 
$\mu^{-} + q \rightarrow e^{-} + q$ (with $q = u,d$) where various 
superparticles run around the loop. In our case with explicit 
$R_p$ violations,
on the contrary, the effective lagrangian Eq.~(13) arises at the tree level 
via superparticle exchanges in different channels.  Therefore, 
the usual loop-induced $\mu^{-} \rightarrow e^{-}$ conversion on the Ti 
nucleus would be suppressed by $O(\alpha / 16 \pi^2 )$ compared with the
tree level contribution from the above effective lagrangian, and thus 
will be neglected in this work.

In order to evaluate the matrix element of the effective lagrangian Eq.~(13) 
between the nucleus as well as the initial and final leptons, we 
assume that the nuclear recoil is negligible,
and the nucleus and the 
initial muon can be treated as nonrelativistic.  Under these assumptions, 
the vector current and the scalar density of the nucleus 
contribute to the coherent conversion process, basically counting the
number of protons and neutrons inside the target nucleus.
Then, the conversion rate for 
the $\mu^{-} + {^{48}_{22}}{\rm Ti} \rightarrow 
e^{-} + {^{48}_{22}}{\rm Ti}$ is given by 
\begin{equation}
\Gamma ( \mu^- + {\rm Ti} \rightarrow e^- + {\rm Ti} ) 
= {\alpha^{3} \over 128 \pi^2}~
{Z_{eff}^4 \over Z}~| F(q^{2} \simeq - m_{\mu}^2 ) |^2 m_{\mu}^5 ~
| Q_{\mu {\rm Ti}}^{eff} |^2,
\end{equation}
where
\begin{eqnarray}
| Q_{\mu {\rm Ti}}^{eff} |^2 & = & 
\left[ (Z+2N) \left( A_{\mu {\rm Ti}}^{d} 
+ S_{\mu {\rm Ti}}^{d,1} + S_{\mu {\rm Ti}}^{d,2}
\right) + A_{\mu {\rm Ti}}^{u} ( 2 Z + N ) \right]^2 
\nonumber    \\
& + & \left[ (Z+2N) \left( A_{\mu {\rm Ti}}^{d} 
+ S_{\mu {\rm Ti}}^{d,1} - S_{\mu {\rm Ti}}^{d,2}
\right) + A_{\mu {\rm Ti}}^{u} ( 2 Z + N ) \right]^2.
\end{eqnarray} 
For ${^{48}_{22}}{\rm Ti}$, one has $Z = 22, N = 26, Z_{eff} = 17.6$ and 
$F(q^{2} \simeq - m_{\mu}^2 ) \simeq 0.54$ \cite{ti}.  

The experimental limit for 
the search for $\mu^- + {_{22}^{48}}{\rm Ti} \rightarrow e^- + 
{_{22}^{48}}{\rm Ti}$ 
is commonly given in terms of 
the above conversion rate divided by
the muon capture rate in ${^{48}_{22}}{\rm  Ti}$, $\Gamma 
( \mu ~{\rm capture ~in 
}~{^{48}_{22}}{\rm Ti} ) = ( 2.590 \pm 0.012 ) \times 10^{6} $/sec 
\cite{pdg} :
\begin{equation}
{\Gamma ( \mu + {\rm Ti} \rightarrow e + {\rm Ti} ) \over 
\Gamma ( \mu ~{\rm capture ~in
}~{^{48}_{22}}{\rm Ti} ) } < 4.3 \times 10^{-12}.
\end{equation}
This puts a strong contraint on $| Q_{\mu {\rm Ti}} |^{2}$ :
\begin{equation}
| Q_{\mu {\rm Ti}}^{eff} |  < 1.2 \times 10^{-9} ~{\rm GeV}^{-2},
\end{equation}
which can be translated into 
\begin{equation}
\left| \left( A_{\mu {\rm Ti}}^{d} + S_{\mu {\rm Ti}}^{d,1}
\pm S_{\mu {\rm Ti}}^{d,2}
\right) + {70 \over 74}~A_{\mu {\rm Ti}}^{u} \right| < 1.6 \times 10^{-7}, 
\end{equation}
for $m_{\tilde{u}_L} = m_{\tilde{d}_R} = m_{\tilde{\nu}_L} = 100 $ GeV.
This is a new strong constraint  which was not considered before
to our knowledge. This is as good as those obtained from $\Delta m_K$,
$\Delta m_{B}$ \cite{roy} or the neutrinoless double beta decay 
experiments \cite{mohapatra}. It also constrains different combinations of 
$R_p-$violating couplings.   



\section{Constraints from $\tau$ decays}
\label{sec:tau}

Now, we consider lepton--family number violating
(LFNV) tau decays into a meson and a lepton, $\tau 
\rightarrow l + PS~ ({\rm or}~ V)$,
where $l=e$ or $\mu$,  $PS = \pi^0, \eta$, 
or $K^{0}$, and $V = \rho^0, \omega, K^{*}$, or $ \phi.$
The relevant effective lagrangian has been already constructed
in the previous subsection, Eqs.~(5) and (7).  
The matrix element for $\langle l, PS({\rm or}~ V) | {\cal L}_{eff} 
| \tau \rangle$ 
can be evaluated using PCAC conditions :
\begin{eqnarray}
\langle \pi^{0} (p) | \overline{u} \gamma_{\mu} \gamma_5 u (0) | 0 \rangle
& = & i f_{\pi} p_{\mu} =  
- \langle \pi^{0} (p) | \overline{d} \gamma_{\mu} \gamma_5 d (0) | 0 \rangle,
\nonumber     
\\
\langle \eta (p)  | \overline{u} \gamma_{\mu} \gamma_5 u (0) | 0 \rangle
& = & {i f_{\pi} \over \sqrt{3}}~ p_{\mu} = 
\langle \eta (p)  | \overline{u} \gamma_{\mu} \gamma_5 u (0) | 0 \rangle
\\
\langle \eta (p)  | \overline{s} \gamma_{\mu} \gamma_5 s (0)
| 0 \rangle & = & -{2 i f_{\pi} \over \sqrt{3}}~ p_{\mu}, 
\nonumber       \\
\langle K (p)  | \overline{d} \gamma_{\mu} \gamma_5 s (0) | 0 \rangle
& = & i \sqrt{2} f_{K}  p_{\mu};
\nonumber  
\end{eqnarray}
and using CVC conditions, 
\begin{eqnarray}
\langle \rho^{0} (p,\epsilon) | \overline{u} \gamma_{\mu} u (0) | 0 \rangle
& = &  m_{\rho} f_{\rho} {\epsilon_{\mu}}^* 
= - \langle \rho^{0} (p,\epsilon) | \overline{d} \gamma_{\mu} d (0)
| 0 \rangle,   \nonumber    \\
\langle \omega^{0} (p,\epsilon) | \overline{u} \gamma_{\mu} u (0) | 0 \rangle
&   = &   m_{\omega} f_{\omega} {\epsilon_{\mu}}^* 
= \langle \omega^{0} (p,\epsilon) | \overline{d} \gamma_{\mu} d (0)
| 0 \rangle,      \\
\langle \phi (p,\epsilon) | \overline{s} \gamma_{\mu} s (0) | 0 \rangle
& = & m_{\phi} f_{\phi} {\epsilon_{\mu}}^*,
\nonumber    \\
\langle K^{*} (p,\epsilon) | \overline{d} \gamma_{\mu} s (0) | 0 \rangle
& = & m_{K^*} f_{K^{*} } {\epsilon_{\mu}}^*.
\nonumber         
\end{eqnarray}
The pseudoscalar meson decay constants, 
$f_{\pi} = 93$ MeV and $f_{K} = 113$ MeV, 
are extracted from the leptonic decay of each pseudoscalar meson,
whereas the vector meson decay constants, 
$f_{\rho} = 153$ MeV,
$f_{\omega} = 138$ MeV, $f_{\phi} = 237$ MeV, and $f_{K^*} = 224$ MeV,
can be obtained from 
$\rho^{0} ({\rm or}~\omega, \phi) \rightarrow e^{+} e^{-}$ and 
$\tau \rightarrow K^{*} + \nu_{\tau}$.

Let us consider $\tau (k,s) \rightarrow e_k (k^{'},s^{'}) + 
V (p, \epsilon)$.   From   
the effective lagrangians Eqs.~(5)-(7), one gets the corresponding amplitude as
\begin{equation}
{\cal M} (\tau \rightarrow e_k + V )  =  {1 \over 8}~
A_{V} f_{V} m_{V} \epsilon_{\mu}^{*}
~\overline{e_k}  \gamma^{\mu} ( 1 - \gamma_{5} ) \tau,
\end{equation}
where 
\begin{eqnarray}
A_{V} & = & A_{V = (u_{j} \overline{u}_{i})} + 
A_{V=(d_{j} \overline{d}_{i})},
\\
A_{V=(u_{j} \overline{u}_{i})} & = & -  \sum_{m,n,p} {V_{mi}^{\dagger} V_{jn}
\over   m_{\tilde{d}_{Rp}}^2}~\lambda_{3mp}^{'} \lambda_{knp}^{'*}
\\
& \rightarrow  & - \sum_{p} { \lambda_{3ip}^{'} \lambda_{kjp}^{'*} \over
m_{\tilde{d}_{Rp}}^2}~~~~~~{\rm for}~K_{np} = \delta_{np}
\\
A_{V=(d_{j} \overline{d}_{i})} & = &  \sum_{m,n,p} {V_{np}^{\dagger} V_{pm}
\over   m_{\tilde{u}_{Lp}}^2}~\lambda_{3nj}^{'} \lambda_{kmi}^{'*}
\\
& \rightarrow  & \sum_{p} { \lambda_{3pi}^{'} \lambda_{kpj}^{'*} \over
m_{\tilde{u}_{Lp}}^2}~~~~~~{\rm for}~K_{np} = \delta_{np}.
\end{eqnarray}
The decay rate for the $\tau \rightarrow e_{k} + V$ is given by 
\begin{equation}
\Gamma ( \tau \rightarrow e_k + V )  =  {1 \over 128 \pi}~|A_{V}|^2 f_{V}^2
~\left[ 2 k \cdot p k^{'} \cdot p + m_{V}^2 k \cdot k^{'} \right]
~{ | \vec{p} | \over  m_{\tau}^2}.
\end{equation}
The limit on the $A_V$ is given in Table~I.   
Note that these limits in Table~I are comparable to
those from $\tau \rightarrow 3e, e \mu^+ \mu^-, 3 \mu$, and so on.
However, these two classes of tau decays constrain different combinations
of $\lambda$ and $\lambda^{'}$ 
from $\tau \rightarrow 3e, e \mu^+ \mu^-,$ or $3 \mu$.   
Therefore, it is worthwhile to consider
$\tau \rightarrow e_k + V$, in addition to $\tau \rightarrow e_k  + \gamma$
and $\tau \rightarrow l l^{'+} l^{'-}$,
as an independent probe of lepton--family number violation beyond SM.
These decays are also easier to study experimentally compared with 
another decays $ \tau \rightarrow e_{k} + PS$ to be considered below, since 
one can tag the dilepton emerging from the decay of a vector meson $V$
(except for $K^{*0}$ which decays mainly into $K \pi$).

Next, consider $\tau (k,s) \rightarrow e_k (k^{'},
s^{'}) + PS (p) $.  There are two contributions : one from the axial vector 
current of quarks, and the other from the pseudoscalar density of quarks.
Using the equations of motion  for the lepton spinors and $p = k - k^{'}$, 
one can transform the former to the latter :
\begin{eqnarray}
p^{\mu} \bar{l} (k^{'},s^{'}) \gamma_{\mu} (1 - \gamma_{5}) \tau (k,s)
& \rightarrow & \bar{l} \left( - m_{l} (1 - \gamma_{5}) + m_{\tau} (1 + 
\gamma_{5} ) \right) \tau     \nonumber 
\\
& \simeq & m_{\tau} \bar{l} ( 1 + \gamma_{5} ) \tau,  
\end{eqnarray}
ignoring the final lepton mass.  
Therefore, the corresponding amplitude derived from the effective
lagrangians, Eqs.~(5) and(6) can be written as 
\begin{equation}
{\cal M} (\tau \rightarrow  e_{k} + PS)  =  
\overline{e_k} ( A_{L}^{PS} P_{L} + A_{R}^{PS} P_{R} ) \tau,
\end{equation}
which leads to the following decay rate :
\begin{equation}
\Gamma ( \tau \rightarrow e_k + PS )  =
{m_{\tau} \over 64 \pi}~\left[  | A_{L}^{PS} + 
A_{R}^{PS} |^2 +  | A_{L}^{PS} - A_{R}^{PS} |^2 \right],
\end{equation}
where $PS (= \pi^0, \eta, K)$  denotes the  final pseudoscalar meson.
We have ignored the final lepton mass compared to the tau mass.
The relevant $A_{L,R}^{PS}$'s   for $PS = \pi^0, \eta, K^{0}$ are given
by the following expressions :
\begin{eqnarray}
A_{L}^{\pi^0} & = & \sum_{n} ~{ \lambda_{n11}^{'*} \lambda_{n3k} 
\over 2 m_{\tilde{\nu}_{Ln}}^2}~{f_{\pi} m_{\pi}^2 \over 2 m_{d}},
\\
A_{R}^{\pi^0} & = & - \sum_{n} ~{ \lambda_{n11}^{'} \lambda_{nk3}^{*}
\over 2 m_{\tilde{\nu}_{Ln}}^2}~{f_{\pi} m_{\pi}^2 \over 2 m_{d}}
 - \sum_{m,n,p}~{ V_{np}^{\dagger}
V_{pm} \over 4 m_{\tilde{u}_{Lp}}^2} ~m_{\tau} f_{\pi} \lambda_{3n1}^{'}
\lambda_{km1}^{'*} 
\nonumber
\\
& & + \sum_{m,n,p}~{V_{m1}^{\dagger} V_{1n} \over 4
m_{\tilde{d}_{Rp}}^2} ~m_{\tau} f_{\pi} ~\lambda_{3mp}^{'}\lambda_{knp}^{'*},
\\
A_{L}^{\eta} & = & - \sum_{n} {\lambda_{n11}^{'*} \lambda_{n3k} \over 2
m_{\tilde{\nu}_{Ln}}^2}~{f_{\pi} m_{\eta}^2 \over \sqrt{3} \times 2 m_{d}}
 + \sum_{n} {\lambda_{n22}^{'*} \lambda_{n3k} \over 2 
m_{\tilde{\nu}_{Ln}}^2}~{2 f_{\pi} m_{\eta}^2 \over \sqrt{3} \times 2 m_{s}},
\\
A_{R}^{\eta} & = & + \sum_{n} {\lambda_{n11}^{'} \lambda_{nk3}^{*} \over 2
m_{\tilde{\nu}_{Ln}}^2}~{f_{\pi} m_{\eta}^2 \over \sqrt{3} \times 2 m_{d}}
 - \sum_{n} {\lambda_{n22}^{'} \lambda_{nk3}^{*} \over 2
m_{\tilde{\nu}_{Ln}}^2}~{2 f_{\pi} m_{\eta}^2 \over \sqrt{3} \times 2 m_{s}}
\nonumber \\
&&  + \sum_{m,n,p} {V_{np}^{\dagger}
V_{pm} \over 4 m_{\tilde{u}_{Lp}}^2}~\left( 
\lambda_{3n1}^{'} \lambda_{km1}^{'*} - 2 \lambda_{3n2}^{'} \lambda_{km2}^{'*}
\right) ~{ f_{\pi} m_{\tau} \over \sqrt{3}}
\\
& & + \sum_{m,n,p} {V_{m1}^{\dagger} V_{1n} \over 4 m_{\tilde{d}_{Rp}}^2}~
\lambda_{3mp}^{'} \lambda_{knp}^{'*}~{ f_{\pi} m_{\tau} \over \sqrt{3}},
\nonumber
\\
A_{L}^{K^0}  & = & - \sum_{n} {\lambda_{n3k} \lambda_{n12}^{'*} \over 2 
m_{\tilde{\nu}_{Ln}}^2}~{ \sqrt{2} f_{K} m_{K}^2 \over (m_{d}+m_{s}) },
\\
A_{R}^{K^0}  & = &  \sum_{n} {\lambda_{nk3}^{*} \lambda_{n21}^{'} \over 2
m_{\tilde{\nu}_{Ln}}^2}~{ \sqrt{2} f_{K} m_{K}^2 \over (m_{d}+m_{s}) }
 + \sum_{m,n,p} {V_{np}^{\dagger} V_{pm}
\over 4 m_{\tilde{u}_{Lp}}^2}~\lambda_{3n1}^{'} \lambda_{km2}^{'*}~\left(
\sqrt{2} f_{K} m_{K} \right).
\end{eqnarray}
In numerical analyses, we use the following current quark masses :
\begin{equation}
m_{u} = 5 ~{\rm MeV}, ~~~~~m_{d} = 10 ~{\rm MeV}, ~~~~~m_{s} = 200~{\rm MeV}.
\end{equation}
Comparing with the experimental upper limits on these SM-forbidden decays,
we get the constraints shown in Table~II.
For the superparticle masses of 100 GeV, the constraints are all order of
$10^{-2} - 10^{-3}$, which are in the similar
range as the constraints obtained from the $\tau \rightarrow e_{k} + V$.
(See Table I.)  


\section{Constraints from  $J/\psi$ and $\pi^0$ decays}
\label{sec:meson}

Finally, let us consider $J/\psi \rightarrow e_{i} \overline{e}_{j}$ 
with $i \neq j$,  and similar decays for $\Upsilon$ and $\pi^0$. 
Since the $J/\psi$ and $\Upsilon$ mainly decay via strong and electromagnetic
interactions, these particles would give weaker constraints on LFNV
couplings compared to the weak transitionis/decays we have considered
before.   However, in these decays, the relevant LFNV couplings 
from the effective lagrangian Eq.~(7) differ from 
those in the others,
and are simpler than those in the  $\tau \rightarrow l + PS$. 
Normalizing the decay rate for the 
$J/\psi \rightarrow e_{i} \overline{e}_{j}$
(with $i \neq j$) to the SM process $J/\psi \rightarrow e^+ e^-$, we get
(summing over two charged modes) 
\begin{equation}
{\Gamma ( J/\psi \rightarrow e_{i} \overline{e_j} + \overline{e_i} e_j )
\over  \Gamma  ( J/\psi \rightarrow e^+ e^- )        }
= {9 \over 64}~{m_{\psi}^4 \over ( 4 \pi \alpha )^2}~|A_{J/\psi}^{(ij)}|^2, 
\end{equation}
with
\begin{equation}
A_{J/\psi}^{(ij)} = \sum_{m,n,p} {V_{m2}^{\dagger} V_{2n} \over
m_{\tilde{d}_{Rp}}^2}~\lambda_{imp}^{'*} \lambda_{jnp}^{'}.
\end{equation}
We have neglected  the final lepton masses.
For the Upsilon decays into $e_i \bar{e}_j$, one can replace 
$m_{\psi}$ by $m_{\Upsilon}$, and multiply the above ratio by a factor of 4
\footnote{Note that $Q_{b} = - |e|/3 = - Q_{c}/2$.}.   

Unfortunately,  there is no published upper limit
on $J/\psi ({\rm or}~\Upsilon (1S)) \rightarrow e \mu, \mu \tau$, or $e \tau$.
For example, the upper limit on the ratio
\begin{equation}
{\Gamma ( J/\psi \rightarrow e^{\pm} \mu^{\mp} ) 
\over \Gamma ( J/\psi \rightarrow 
e^{+} e^{-} ) } < 10^{-4}
\end{equation}
would imply  $| A_{J/\psi}^{(12)} | < 7.2$ for 
$m_{\tilde{d}_{R}} = 100$ GeV.
As one might expect, this limit is not that stringent, since $J/\psi$
(and $\Upsilon$) decays mainly through strong and electromagnetic 
annihilations, and not through weak annihilation.
However, one may still try to search for
the LFNV $J/\psi$ decays at Tau-Charm factories.
Note that $\lambda_{22p}^{'} \lambda_{12p}^{'}$ 
has never been constrained before.

Similarly, the effective lagrangians, Eqs.~(5) and (7) 
contribute to the  decays 
$\pi^{0} \rightarrow e^{+} e^{-}$  and $\eta \rightarrow l^{+} l^{-}$ as 
well as the LFNV decay $\pi^0  \rightarrow e^{\pm} \mu^{\mp}$.
In these decays, the (pseudo)scalar $\times$ (pseudo)scalar couplings in 
Eq.~(5) give the largest contributions because they are enhanced by
a factor of $m_{\pi}^2/ m_{d}$ compared with $m_{\pi}$ or $m_{\mu}$, 
if the couplings and the masses of the superparticles are in the
same order of magnitude.  So we ignore the contributions from $(V-A)$ quark
currents in Eqs.~(5)-(7).  
In this approximation, the amplitude for $\pi^0 \rightarrow e_i \bar{e}_j$
becomes 
\begin{equation}
{\cal M} (\pi^0 \rightarrow e_i \bar{e}_j ) 
= A_{P,L} ~\overline{e_{j,L}} e_{i,R}~ + ~A_{P,R} ~\overline{e_{j,R}}
e_{i,L},
\end{equation}
with
\begin{eqnarray}
 A_{P,L} = {f_{\pi} m_{\pi}^{2} \over 8 m_{d}}~\sum_{n} {\lambda_{n11}^{'}
\lambda_{nij}^{*} \over m_{\tilde{\nu}_{Ln}}^2},
\\
 A_{P,R} = - {f_{\pi} m_{\pi}^{2} \over 8 m_{d}}~\sum_{n} {\lambda_{n11}^{'*}
\lambda_{nji} \over m_{\tilde{\nu}_{Ln}}^2}.
\end{eqnarray}
The resulting decay rate is (after summing over the charge conjugate state)
\begin{equation}
\Gamma ( \pi^0 \rightarrow \mu^{\pm} e^{\mp} )
= {m_{\pi} \over 16 \pi}~\left[
 | A_{P,L} + A_{P,R} |^2 +  | A_{P,L} - A_{P,R} |^2 \right]~
\left( 1 - {m_{\mu}^2 \over m_{\pi}^2} \right)^2.
\end{equation}
For the LFNV decays $\pi^0 \rightarrow e^{\pm} \mu^{\mp}$, there is a tight 
upper bound on the branching ratio, $ 1.72 \times 10^{-8}$.    
This implies that (for $m_{\tilde{\nu_{Ln} }} = 100$ GeV) 
\begin{equation}
\left| \sum_{n} { { \left( \lambda_{n11}^{'}\lambda_{n12}^{*} \pm 
\lambda_{n11}^{'*}\lambda_{n21} \right) }
} \right| < 0.14.
\end{equation}
For the lepton number conserving decay $\pi^0 \rightarrow e^+ e^-$,
the branching ratio is known to be
\begin{equation}
B (\pi^0 \rightarrow e^+ e^- ) = (7.5 \pm 2.0) \times 10^{-8},
\end{equation}
which is dominated by the so-called unitarity bound coming from 
$\pi^0 \rightarrow \gamma\gamma \rightarrow e^+ e^-$. This unitarity bound
is calculable, and known to be \cite{ko}
\begin{equation}
{{ \Gamma_{unit} (\pi^0 \rightarrow e^+ e^-) } \over 
{ \Gamma (\pi^{0} \rightarrow \gamma \gamma ) }} =  {\alpha^2 \over 2 
\beta_{e}}~{m_e^2 \over m_{\pi}^2}~\left[ \ln \left( {{1+\beta_e} \over 
{1-\beta_e}} \right) \right]^2 = 4.75 \times 10^{-8},
\end{equation}
with $\beta_{e} = \sqrt{1- 4m_{e}^2 / m_{\pi^0}^2}$.  
Extracting this unitary bound from the experimental branching ratio and 
assuming no large contributions from the dispersive part of the two-photon
contributions ($ {\rm Re} A_{\gamma \gamma}$) or large cancellation between
$ {\rm Re} A_{\gamma \gamma}$ and the $R_p-$violating contributions, 
we can put the  (90 \% C.L.)  limit
on the contribution from the $R_p-$violating interactions in Eq.~(5) :
\begin{equation}
\left| \sum_{n} {{ \left( \lambda_{n11}^{'}\lambda_{n11}^{*} \pm
\lambda_{n11}^{'*}\lambda_{n11} \right) }
}  \right| < 0.15,
\end{equation}
for $m_{\tilde{\nu}_L} = 100$  GeV.


\section{Conclusions}
\label{sec:con}

In conclusion, we  considered several  different LFNV processes : 
(i) the muonium conversion,
(ii) $\mu^- + {_{22}^{48}}{\rm Ti} \rightarrow e^- + {_{22}^{48}}{\rm Ti}$,
(iii) $\tau$ decays into a lepton and a meson, $\tau \rightarrow 
l + PS~({\rm or}~V)$, and
(iv) $J/\psi (\Upsilon, \pi^{0}) \rightarrow e_{i}
\overline{e_k}$. From these processes, we got constraints on the 
$R-$parity violating couplings and the superparticle masses. 
Some of these constraints are new, and/or stronger than constraints 
from other popular processes such as 
$\tau \rightarrow \mu (e) + \gamma$, $\tau \rightarrow l l^{' +} l^{' -}$,
{\it etc.}.  We got one of the strongest constraints, Eq.~(22), 
from $\mu^{-}
\rightarrow e^{-}$ conversion on the ${^{48}_{22}}$Ti nucleus.  
This originates from the fact that the $R-$parity violating terms can give
tree-level contributions to the processes considered in this work. 
In many supersymmetric models with $R-$parity conservation, 
on the other hand,
LFNV processes usually arise from one-loop Feynman diagrams, so that the
most important one is often the electromagnetic penguin
($\mu^- \rightarrow e^- \gamma$) contribution to 
$\mu^- + {^{48}_{22}}{\rm Ti}
\rightarrow e^-  +  {^{48}_{22}}{\rm Ti}$.
Therefore, dedicated searches for these  decays at PSI, Tau--Charm factories
and other facilities are clearly warranted, and are very important,
because
they will provide us with hints of new physics beyond the SM via lepton 
flavor violations from any origin, including those from supersymmetric
models with/without $R-$parity conservations.

\vspace{.3in}

\acknowledgements

P.K. is grateful to Dr. E.J. Chun for discussions on the $R-$parity 
violations.
This work was supported in part by KOSEF through CTP at Seoul National 
University, by the Basic Science Research Program, Ministry of 
Education, 1996, Project No. BSRI--96--2418, Korea-Nagoya Exchange
Program, and by the Distinguished Scholar Exchange Program of Korea 
Research Foundation.


%
%

%
%
\begin{table}
\caption{Constraints from $\tau \rightarrow l + V$ with $l=e$ or $\mu$, 
and $V = \rho^{0}, K^{*}$ or $\phi$.
In the table we use the notations,
$u_{p} \equiv ( 100~{\rm GeV} / m_{\tilde{u}_{Lp}} )^2$ and
$d_{p} \equiv ( 100~{\rm GeV} / m_{\tilde{d}_{Rp}} )^2$.
Data are taken from the recent results 
reported by CLEO Collaborations, Ref.~[17].
Sum over $m,n,p=1,2,3$ is to be  understood.}
\label{table1}
\begin{tabular}{cccc}
final state & $B_{exp}$ & Combinations constrined & Constraint
\\
\tableline
$e \rho^0$ & $ < 4.2 \times 10^{-6}$ &   
$  V_{np}^{\dagger} V_{pm} \lambda_{3n1}^{'} \lambda_{1m1}^{'*} 
u_{p}$
& $ < 3.5 \times 10^{-3}$  
\\
 & &
$V_{n1}^{\dagger} V_{1m} \lambda_{3np}^{'} \lambda_{1mp}^{'*} d_{p}$
& $ < 3.5  \times 10^{-3}$
\\
$e K^{*0}$ & $ < 6.3 \times 10^{-6}$ &
$  V_{np}^{\dagger} V_{pm} \lambda_{3n1}^{'} \lambda_{1m2}^{'*} 
u_{p}$
& $ < 3.0 \times 10^{-3}$ 
\\
$\mu \rho^{0}$ & $ < 5.7 \times 10^{-6}$ &
$  V_{np}^{\dagger} V_{pm} \lambda_{3n1}^{'} \lambda_{2m1}^{'*} 
u_{p} $
& $ < 4.2 \times 10^{-3}$ 
\\
 & &
$V_{n1}^{\dagger} V_{1m} \lambda_{3np}^{'} \lambda_{2mp}^{'*} d_{p}$
& $ < 4.2  \times 10^{-3}$
\\
$\mu K^{*0}$ & $ < 9.4  \times 10^{-6}$ &
$  V_{np}^{\dagger} V_{pm} \lambda_{3n1}^{'} \lambda_{2m2}^{'*} 
u_{p} $
& $ < 3.8 \times 10^{-3}$ 
\end{tabular}
\end{table}

\begin{table}
\caption{Constraints from $\tau \rightarrow l + PS$ with $l=e$ or $\mu$,
and $PS$ being a light pseudoscalar meson. 
In the table we use the notations,
$n_{n} \equiv ( 100~{\rm GeV} / m_{\tilde{\nu}_{Ln}} )^2$,
$u_{p} \equiv ( 100~{\rm GeV} / m_{\tilde{u}_{Lp}} )^2$, and
$d_{p} \equiv ( 100~{\rm GeV} / m_{\tilde{d}_{Rp}} )^2$.
Data are taken from Ref.~[16].
Sum over $m,n,p$ is to be understood.  }
\label{table2}
\begin{tabular}{cccc}
final state & $B_{exp}$ & Combinations constrained & Constraint
\\
\tableline
$e \pi^0$ &  $ < 1.4 \times 10^{-4} $  & 
$ \lambda_{n31} \lambda_{n11}^{'*} n_{n},  
\lambda_{n13}^{*} \lambda_{n11}^{'} n_{n} $
& $ < 6.4 \times 10^{-2}$
\\
         &  & 
$V_{np}^{\dagger} V_{pm} \lambda_{3n1}^{'} \lambda_{1m1}^{'*} u_{p},
V_{m1}^{\dagger} V_{n1} \lambda_{3mp}^{'} \lambda_{1np}^{'*} d_{p}$
& $ < 6.6 \times 10^{-2}$
\\
$\mu \pi^0$ & $ < 4.4 \times 10^{-5} $ &
$ \lambda_{n32} \lambda_{n11}^{'*} n_{n},  
\lambda_{n23}^{*} \lambda_{n11}^{'} n_{n} $
& $ < 3.6 \times 10^{-2}$ 
\\
         &  &
$V_{np}^{\dagger} V_{pm} \lambda_{3n1}^{'} \lambda_{2m1}^{'*} u_{p},
V_{m1}^{\dagger} V_{n1} \lambda_{3mp}^{'} \lambda_{2np}^{'*} d_{p}$
& $ < 3.7 \times 10^{-2}$
\\
$e K^0$   &  $ < 1.3 \times 10^{-3} $  &
$ \lambda_{n31} \lambda_{n12}^{'*} n_{n},
\lambda_{n13}^{*} \lambda_{n21}^{'} n_{n} $
& $ < 8.5  \times 10^{-2}$
\\
 & & 
$V_{np}^{\dagger} V_{pm} \lambda_{3n1}^{'} \lambda_{1m2}^{'*} u_{p}$
& $ < 4.0  \times 10^{-1}$
\\
$\mu K^0$ & $ < 1.0 \times 10^{-3}$ &
$ \lambda_{n32} \lambda_{n12}^{'*} n_{n},
\lambda_{n23}^{*} \lambda_{n21}^{'} n_{n} $
& $ < 7.6  \times 10^{-2}$
\\
 & & 
$V_{np}^{\dagger} V_{pm} \lambda_{3n1}^{'} \lambda_{2m2}^{'*} u_{p}$
& $ < 3.6  \times 10^{-1}$
\\
$e \eta$ & $ < 6.3 \times 10^{-5} $ &
$ \lambda_{n31} \lambda_{n11}^{'*} n_{n},
\lambda_{n13}^{*} \lambda_{n11}^{'} n_{n} $
& $ < 4.5  \times 10^{-3}$
\\
  &   &   
$\lambda_{n31} \lambda_{n22}^{'*} n_{n},
\lambda_{n13}^{*} \lambda_{n22}^{'} n_{n} $
& $ < 4.5  \times 10^{-2}$
\\
   & &
$V_{np}^{\dagger} V_{pm} ( \lambda_{3n1}^{'} \lambda_{1m1}^{'*} - 
2 \lambda_{3n2}^{'} \lambda_{1m2}^{'*} ) u_{p}$ & $ < 7.8 \times 10^{-2} $ 
\\    & & 
$V_{m1}^{\dagger} V_{1n} \lambda_{3mp}^{\dagger} \lambda_{1np}^{'*} d_{p}$
& $  < 7.8 \times 10^{-2} $ 
\\
$\mu \eta $ & $ < 7.3 \times 10^{-5} $ &
$ \lambda_{n32} \lambda_{n11}^{'*} n_{n},
\lambda_{n23}^{*} \lambda_{n11}^{'} n_{n} $
& $ < 4.8  \times 10^{-3}$
\\
  &   &
$\lambda_{n32} \lambda_{n22}^{'*} n_{n},
\lambda_{n23}^{*} \lambda_{n22}^{'} n_{n} $
& $ < 4.8  \times 10^{-2}$
\\
   & &
$V_{np}^{\dagger} V_{pm} ( \lambda_{3n1}^{'} \lambda_{2m1}^{'*} -
2 \lambda_{3n2}^{'} \lambda_{2m2}^{'*} ) u_{p}$   &  $ < 8.2 \times 10^{-2} $
\\   &  &   
$V_{m1}^{\dagger} V_{1n} \lambda_{3mp}^{\dagger} \lambda_{2np}^{'*} d_{p}$
& $  < 8.2 \times 10^{-2} $
\end{tabular}
\end{table}

\end{document}